\documentstyle[12pt]{article}
\def\bge{\begin{equation}}
\def\ene{\end{equation}}
\def\bg{\begin{eqnarray}}
\def\en{\end{eqnarray}}

\def\bi{\bibitem}
\begin{document}
\begin{flushright}
ADP-95-32/T186
\end{flushright}
\begin{center}
\begin{Large}
\center{Composite nucleons in scalar and vector mean-fields}
\end{Large}
\end{center}
\vspace{0.3cm}
\begin{center}
\begin{large}
K.~Saito~\footnote{ksaito@nucl.phys.tohoku.ac.jp} \\
Physics Division, Tohoku College of Pharmacy \\
Sendai 981, Japan \\
and \\
A.~W.~Thomas~\footnote{athomas@physics.adelaide.edu.au} \\
Department of Physics and Mathematical Physics \\
and \\
Institute for Theoretical Physics, \\
University of Adelaide, South Australia 5005, Australia
\end{large}
\end{center}
\vspace{0.3cm}
\begin{abstract}
We emphasize that the composite structure of the nucleon
may play quite an important role in nuclear physics.
It is shown that the momentum-dependent repulsive force
of second order in the scalar field, which plays an important role in
Dirac phenomenology, can be found in the quark-meson coupling (QMC) model,
and that the properties of nuclear matter are well described through
the quark-scalar density in a nucleon and a self-consistency condition
for the scalar field. The difference between
theories of point-like nucleons and composite ones may be seen in the
change of the $\omega$-meson mass in nuclear matter if the composite nature
of the nucleon suppresses contributions from nucleon-antinucleon pair
creation.
\end{abstract}
%
%%%%%%%%%%%%%%%%%%%%%%%%%%%%%%%%%%%%%%%%%%%%%%%%%%%%%%%%%%%%%%%%%%%%%
%
\newpage

It is well known that relativistic theories of nucleons interacting with
mesons are very powerful in the treatment of a wide
range of nuclear phenomena (Dirac phenomenology), most notably the
single particle energy levels, nuclear charge densities
and elastic proton-nucleus scattering
observables at intermediate energies~\cite{qhd,ann,rpp}.
The simplest and earliest example is
the $\sigma$-$\omega$ model of Walecka~\cite{wal} (sometimes called
QHD~\cite{qhd}), which consists of
{\em structureless} nucleons interacting with each other through the
exchange of the scalar ($\sigma$) and the vector ($\omega$) mesons.

These typically involve large scalar ($S$) and vector ($V$) potentials
of opposite
sign, which provide a number of interesting effects -- e.g. a strong
momentum-dependence of the optical potential and an enhanced spin-orbit
force~\cite{qhd,ann,rpp,jam}.  One approach to understanding the
physics content of Dirac
phenomenology is to emphasise the role of {\em virtual nucleon-antinucleon}
($N{\bar N}$) pair creation.
A simple estimate, up
to second order
in the scalar field, shows a potential
which contains the effect of couplings to virtual $N{\bar N}$-pair
states~\cite{ann}:
\bg
U_{pair} \approx \frac{{\bf \sigma}\cdot{\bf p}}{2M_N}
\frac{(S-V)^2}{2M_N} \frac{{\bf \sigma}\cdot{\bf p}}{2M_N}
\approx \frac{{\bf p}^2}{2M_N^3}S^2  , \label{upair}
\en
if $V \approx -S$.  This repulsive, strongly momentum-dependent term
plays an important role in producing nuclear saturation
and in enhancing
the spin-orbit coupling in Dirac phenomenology.  From this point of
view, the
excitation of virtual $N{\bar N}$ pairs (i.e., $Z$-graphs), is a vital
ingredient in the success of this approach.

However, some people have criticized the idea that $N{\bar N}$ creation
should play such an important role.
Brodsky~\cite{brod}
has argued that the pair creation should be supressed by form factors for
composite objects.  Kiritsis and Seki~\cite{ks} have shown
that baryon
loops are supressed in the $1/N_c$ expansion of QCD.  Using some tractable
models,
Cohen~\cite{cohen} has also emphasized that the composite nature of the
nucleon supresses the contribution of $N{\bar N}$ pairs compared with what
is expected in the naive Dirac phenomenology.  Furthermore, Prakash
et al.~\cite{prak} have shown that the composite structure of the
nucleon ought to largely soften the 2-loop contributions~\cite{2loop}
in QHD.  Then, is Dirac phenomenology in doubt ?

Recently Wallace, Gross and Tjon~\cite{wall} have pointed out that scalar
and vector interactions, which couple to a composite spin-1/2 system,
obey a low-energy theorem which guarantees the same
repulsive second-order interaction as given in Eq.(\ref{upair}).
Later Birse~\cite{birse} discussed it in a more general
fashion, and showed, without referring to any nucleon $Z$-graph,
that not only can there be a
momentum-dependent, repulsive force (as in Eq.(\ref{upair})),
but one may also find other types of second-order interaction
which depend on the
nucleon structure through various polarisabilities.  It is
known that in the case of the soft-photon limit of Compton
scattering~\cite{jen} and low-energy theorems for $\pi$-N
interactions~\cite{birse2} quark excitations and {\em quark}
$Z$-graphs conspire
to produce the same results as nucleon $Z$-graphs.  As Cohen
has noticed~\cite{cohen}, Dirac phenomenology depends
only on the presence of strong scalar and vector potentials in the
{\em effective} one-body optical potential, and there is no logical need
for such forms to be associated with the excitation of $N{\bar N}$ pairs.

The momentum-dependent, repulsive interaction
can be also seen
in the quark-meson coupling (QMC) model~\cite{guichon,st1}.
In this model the properties of nuclear matter
are determined by the self-consistent coupling of scalar and
vector mean-fields
 to the {\it quarks}, rather than the nucleons.
In a simple model,
where nuclear matter was considered as a collection of
static, non-overlapping bags, it was shown that a satisfactory
description of the bulk properties of nuclear matter could be
obtained.
Furthermore, the model seems
to provide a semi-quantitative explanation of the Okamoto-Nolen-Schiffer
anomaly~\cite{ok} when quark mass differences are included~\cite{st2},
as well as the nuclear EMC effect~\cite{st3}.

In the QMC model the energy of a nucleon with momentum
$\bf p$ interacting with both $\sigma$ and $\omega$ mean-fields in
the rest frame of uniform nuclear matter is given by
\bg
E({\bf p}) = g_{\omega} {\bar \omega} + \sqrt{{\bf p}^2 +
M_N^{\star}({\bar \sigma})^2}  , \label{energy}
\en
where ${\bar \sigma}$ and ${\bar \omega}$ are the mean-field values,
$M_N^{\star}$ is the effective nucleon mass, which is a function of
${\bar \sigma}$, and the vector field couples to the conserved
baryon current with strength $g_{\omega}$.
At low nuclear density $M_N^{\star}$
can be expanded in terms of the scalar field as
\bg
M_N^{\star}({\bar \sigma}) = M_N + \left( \frac{d M_N^{\star}}
{d {\bar \sigma}}
\right)_{{\bar \sigma}=0} {\bar \sigma} + \frac{1}{2} \alpha_s
{\bar \sigma}^2 + \cdots  , \label{nmass}
\en
where
$M_N$ is the free nucleon mass and
$\alpha_s$ is the second derivative of $M_N^{\star}$ with respect to
${\bar \sigma}$.
We can easily see
that the second term on the r.h.s. of
Eq.(\ref{nmass}) is a response function to the external
scalar field, and that it is given by the scalar density of a quark in the
nucleon bag:
\bg
\left( \frac{d M_N^{\star}}{d {\bar \sigma}} \right) \equiv
-g_{\sigma} C_N(\bar{\sigma})
= -g_{\sigma} \int_{R_N} d{\vec r} {\bar \psi}_q \psi_q  . \label{cn}
\en
Here $g_{\sigma}$ is the coupling constant of the $\sigma$ field
to the nucleon.
(If a correction for spurious center of mass motion in the bag is
taken into account~\cite{yaz}, the r.h.s. of Eq.(\ref{cn}) is modified
accordingly~\cite{st1}.)
Therefore, since
\bg
M_N^{\star} \simeq M_N - g_{\sigma} C_N(0) {\bar \sigma} + \frac{1}{2}
\alpha_s {\bar \sigma}^2 , \label{nmass2}
\en
we find the nucleon energy up to ${\cal O}({\bar \sigma}^2)$ as
\bg
E({\bf p}) \simeq g_{\omega} {\bar \omega} + \epsilon({\bf p})
- g_{\sigma} \frac{C_N(0) M_N}{\epsilon({\bf p})} {\bar \sigma}
+ \frac{\alpha_s M_N}{2 \epsilon({\bf p})} {\bar \sigma}^2
+ g_{\sigma}^2 \frac{C_N(0)^2 {\bf p}^2}{2 \epsilon({\bf p})^3}
{\bar \sigma}^2  , \label{energy2}
\en
where $\epsilon({\bf p}) = \sqrt{ {\bf p}^2 + M_N^{\star 2}}$.  If we
replace $g_{\sigma} C_N(0) {\bar \sigma}$ by the
scalar potential $S$, the last
term in Eq.(\ref{energy2}) is indeed the momentum-dependent repulsive
force pointed out by Wallace et al.~\cite{wall} and
Birse~\cite{birse} (see Eq.(\ref{upair})).
One can see that such a term appears in any
relativistic treatment and that
it arises from the
modification of the nucleon mass due to the scalar field.

In our model the effect of the
internal,
quark structure of a nucleon can be completely absorbed into the
scalar density, $C_N({\bar \sigma})$.  The self-consistency condition (SCC)
for the $\sigma$ field is then given by
\bg
g_{\sigma}\bar{\sigma} = \frac{g_{\sigma}^2}{m_{\sigma}^2}\frac{\gamma}
{(2\pi)^3}C_N(\bar{\sigma}) \int^{p_F} d{\bf p}
\frac{M_N^{\star}}{\sqrt{M_N^{\star 2} + {\bf p}^2}} , \label{eq:scc}
\en
where $\gamma$ is the spin-isospin degeneracy factor, $m_{\sigma}$ is the
mass of the $\sigma$ and $p_F$ is the Fermi
momentum for the nucleon.  If we set
$C_N = 1$, the above SCC becomes identical to that of QHD~\cite{st1}.
Therefore, it is quite important to examine this scalar density
in order to understand the
difference between theories of point-like nucleons and composite ones.
In Fig.\ref{fig:scald} $C_N$ is
shown as a function of the nuclear density, $\rho_B$. (The normal nuclear
density is denoted by $\rho_0$, and the coupling constants
have been chosen to reproduce the nuclear saturation properties~\cite{st1}.)
%
%\begin{figure}[tb]
%\begin{center}
%\epsfile{file=comp1.eps,height=9cm}
%\epsfig{file=comp1.eps,height=9cm}
%\caption{Quark-scalar density for various bag radii ($R_0$) as a function
%of $\rho_B$.
%The solid, dotted and dashed
%curves show $C_N$ for $R_0$ = 0.6, 0.8 and 1.0 fm,
%respectively.
%The quark mass is chosen to be 5 MeV.}
%\label{fig:scald}
%\end{center}
%\end{figure}
%
\begin{center}
\fbox{Figure~\ref{fig:scald}}
\end{center}

Clearly the scalar density, $C_N(\bar{\sigma})$, is much less than
unity, and depends strongly on the nuclear density -- as $\rho_B$
goes higher $C_N$ becomes smaller.  This is because
the small component of the quark wave function responds rapidly to the
scalar field.
As the scalar density itself is the source of the
$\sigma$ field this provides a suppression of the $\sigma$ field at
high density, and hence a new mechanism for the saturation of nuclear
matter where the quark structure plays a vital role.
Of particular interest is the fact that the internal structure of
the nucleon results in a
lower value of the incompressibility of nuclear
matter than that obtained in
approaches based on point-like nucleons -- e.g. as in QHD~\cite{qhd}.
In fact, our prediction ($\sim$ 220 MeV)~\cite{st1} is in
agreement with the experimental value once the binding energy and
saturation density are fixed.   The effect of the quark-structure
of the nucleon on the spin-orbit force in finite nuclei has been discussed
in Ref.\cite{new}.

One of the most topical questions which can be addressed within this
model is the change of hadron properties in matter.
In particular,
variations in hadron masses have attracted wide
%interest~\cite{hatkun,vmd,qcdsum,qhd2,qhd3,st4}.
interest [22-27].
It is therefore very interesting to compare the prediction of the
$\omega$-meson mass in matter by QHD~\cite{qhd2,qhd3} with that by
the QMC model~\cite{st4}.  In the latter, if we suppose that the $\omega$
meson is also described by the MIT bag model in the scalar mean-field, the
effective $\omega$-meson mass, $m_{\omega}^{\star}$,
at low density is given
(as in the nucleon case) by:
\bg
m_{\omega}^{\star} \simeq m_{\omega} - \frac{2}{3} g_{\sigma} C_{\omega}(0)
{\bar \sigma}  , \label{omass}
\en
where $m_{\omega}$ is the free mass of $\omega$ and
$C_{\omega}({\bar \sigma})$ is the quark-scalar density in the $\omega$
meson.  Eq.(\ref{omass})
means that the $\omega$-meson mass in matter decreases as $\rho_B$
grows: $(m_{\omega}^{\star}/m_{\omega}) \simeq 1 -
0.09 (\rho_B/\rho_0)$~\cite{st4}.  On the other hand, in QHD
the $\omega$-meson mass at low density is given by~\cite{qhd3}
\bg
m_{\omega}^{\star} \simeq m_{\omega} + \frac{1}{2} \frac{\Omega^2}
{m_{\omega}} - \frac{g_{\sigma}^2 m_{\omega} \Omega^2}{6 \pi^2
m_{\sigma}^2}  , \label{omass2}
\en
where $\Omega^2 = g_{\omega}^2 \rho_B / M_N$ is the classical plasma
frequency.  The second term on the r.h.s of Eq.(\ref{omass2}) comes from
the density-dependent part of the
$\omega$-meson propagator in random-phase approximation, which,
as reported by Chin~\cite{chin}, leads to an {\em increase} in the mass.
The third term, which gives a strong, attractive contribution,
is due to vacuum polarisation. Finally, the sum of both
of these effects gives a decrease of the
mass.  In QHD the contribution of vacuum polarisation, i.e.,
$N{\bar N}$ pair creation, is essential~\cite{qhd3} to reproduce the mass
reduction predicted by the QCD sum rules~\cite{qcdsum,qcdsum2}.

We emphasise that the origins of the mass reduction in QHD and the QMC model
are completely different.  As noticed by some
people~\cite{brod,ks,cohen},
if the contribution from vacuum polarization were largely suppressed
in QHD the $\omega$ mass would be given mainly by the first and
second terms of Eq.(\ref{omass2}) and would increase in matter.  In fact, it is
proven that vertex corrections are quite important in QHD and that
such corrections dramatically reduce the vacuum
contributions in comparison with those calculated with bare vertices
in 2-loop calculations~\cite{prak,2loop}.  This means that in the
2-loop case the nucleons are dressed with meson clouds or, more
generally, they have {\em structure}, and that this compositeness
suppresses vacuum contributions from $N{\bar N}$ loops.

In conclusion, we have argued that the composite structure of the nucleon
may play quite an important role in nuclear physics.
The momentum-dependent repulsive force
of second order in the scalar field, which plays an important role in
Dirac phenomenology, can be found in any relativistic model of
composite nucleons involving scalar and vector mean-fields.  In the QMC model
the properties of
nuclear matter can be well reproduced through the quark-scalar
density in the nucleon and the self-consistency condition
for the scalar field.  We have pointed out that
theories of point-like nucleons may be
distinguishable from those involving the internal structure of the
nucleon (and other hadrons) through the
behaviour of the $\omega$-meson mass in matter.
In particular, if $N{\bar N}$ pair creation were strongly suppressed,
one might even find an increase of the $\omega$-meson mass.
This is quite the opposite of the behaviour found in models of
composite nucleons, such as the QMC model.
Clearly it would be extremely valuable to have some experimental
guidance on this matter.

\vspace{1cm}
Acknowledgements

We would like to acknowledge helpful discussions with A.~G.~Williams.
This work was supported by the Australian Research Council.

%%%%%%%%%%%% Bibliography %%%%%%%%%%%%%%%%%%%%%%%%%%%%%%%%%
%

%
\newpage
\begin{flushleft}
\begin{Large}
\underline{Figure caption}
\end{Large}
\end{flushleft}

\noindent Figure 1; Quark-scalar density for various bag radii ($R_0$)
as a function of $\rho_B$.  The solid, dotted and dashed
curves show $C_N$ for $R_0$ = 0.6, 0.8 and 1.0 fm,
respectively.  The quark mass is chosen to be 5 MeV.
\newpage
\begin{figure}
\begin{center}
%\epsfile{file=comp1.eps,height=15cm}
%\epsfig{file=comp1.eps,height=15cm}
\caption{ }
\label{fig:scald}
\end{center}
\end{figure}

\end{document}